\documentclass{article} % For LaTeX2e
\pdfoutput=1
\usepackage{preiclr,times}
\usepackage{hyperref}
\usepackage{url}
\usepackage{amssymb}
\usepackage{amsfonts}
\usepackage{graphicx}

\newcommand{\expectedvalue}{\mathbb{E}}
\newcommand{\argmin}{\mathrm{argmin}}

\title{Learning to Protect Communications\\ with Adversarial Neural Cryptography}
\author{Mart{\'\i}n Abadi and David G. Andersen \thanks{Visiting from Carnegie Mellon University.}\\
Google Brain}

\begin{document}
\maketitle
\begin{abstract}
We ask whether neural networks can learn to use secret keys to protect
information from other neural networks.  Specifically, we focus on
ensuring confidentiality properties in a multiagent system, and we
specify those properties in terms of an adversary.  Thus, a
system may consist of neural networks named Alice and Bob, and we aim
to limit what a third neural network named Eve learns from
eavesdropping on the communication between Alice and Bob.
We do not prescribe specific cryptographic algorithms to these neural networks;
instead, we train end-to-end, adversarially.
We demonstrate that the neural networks can learn 
how to perform forms of encryption and decryption, and also
how to apply these operations selectively in order to meet
confidentiality goals.
\end{abstract}

\section{Introduction}

As neural networks are applied to increasingly complex tasks, they are
often trained to meet end-to-end objectives that go beyond simple
functional specifications. These objectives include, for example,
generating realistic images
(e.g.,~\citep{DBLP:conf/nips/GoodfellowPMXWOCB14}) and solving
multiagent problems
% that require the discovery of communication protocols
(e.g.,~\citep{DBLP:journals/corr/FoersterAFW16,DBLP:journals/corr/FoersterAFW16a,DBLP:journals/corr/SukhbaatarSF16}).
Advancing these lines of work, we show that neural networks can learn
to protect their communications in order to satisfy a policy specified in
terms of an adversary.

Cryptography is broadly concerned with algorithms and protocols that
ensure the secrecy and integrity of information.
% (e.g.,~\citep{Menezes:1996:HAC:548089}).
Cryptographic
mechanisms are typically described as programs or Turing machines. 
Attackers
are also described in those terms, with bounds on their complexity
(e.g., limited to polynomial time) and on their chances of success
(e.g., limited to a negligible probability). A mechanism is deemed
secure if it achieves its goal against all attackers. For instance, an
encryption algorithm is said to be secure if no attacker can extract information
about plaintexts from ciphertexts. Modern cryptography provides
rigorous versions of such
definitions~\citep{DBLP:journals/jcss/GoldwasserM84}.

Adversaries also play important roles in the design and training
of neural networks. They arise, in particular, in work on adversarial
examples~\citep{DBLP:journals/corr/SzegedyZSBEGF13,DBLP:journals/corr/GoodfellowSS14} and on generative
adversarial networks
(GANs)~\citep{DBLP:conf/nips/GoodfellowPMXWOCB14}. In this latter context, 
the adversaries
are neural networks (rather than Turing machines) that attempt to determine whether a sample value
was generated by a model or drawn from a given data distribution. Furthermore, in contrast with definitions in cryptography,
practical approaches to training GANs do not consider all possible adversaries in a class, but
rather one or a small number of adversaries that are optimized by
training. We build on these ideas in our work.

Neural networks are generally not meant to be great at
cryptography. Famously, the simplest neural networks cannot
even compute XOR,
% ~\citep{Minsky:1988:PEE:50066},
which is basic to many cryptographic algorithms.  Nevertheless, as we
demonstrate, neural networks can learn to protect the confidentiality
of their data from other neural networks: they discover
forms of encryption and decryption, without being taught specific algorithms for these purposes.

Knowing how to encrypt is seldom enough for security and privacy.
Interestingly, neural networks can also learn \emph{what} to encrypt in order
to achieve a desired secrecy property while maximizing utility. Thus, when we wish to prevent
an adversary from seeing a fragment of a plaintext, or from estimating
a function of the plaintext, encryption can be
selective, hiding the plaintext only partly.

The resulting cryptosystems are generated automatically. In this
respect, our work resembles recent research on automatic synthesis of
cryptosystems, with tools such as
ZooCrypt~\citep{Barthe:2013:FAA:2541806.2516663}, and contrasts with
most of the literature, where hand-crafted cryptosystems are the norm.
ZooCrypt relies on symbolic theorem-proving, rather than neural
networks.

Classical cryptography, and tools such as ZooCrypt, typically
provide a higher level of transparency and assurance than we would
expect by our methods.  Our model of the adversary, which avoids
quantification, results in much weaker guarantees. On the other hand, it is
refreshingly simple, and it may sometimes be appropriate.

Consider, for example, a neural network with several components, and
suppose that we wish to guarantee that one of the components does not
rely on some aspect of the input data, perhaps because of concerns
about privacy or discrimination. Neural networks are notoriously
difficult to explain, so it may be hard to characterize how the
component functions.  A simple solution is to treat the component as
an adversary, and to apply encryption so that it does not have access
to the information that it should not use. 
% We explore such applications below.
In this respect, the
present work follows the recent research on fair
representations~\citep{DBLP:journals/corr/EdwardsS15,DBLP:journals/corr/LouizosSLWZ15}, 
which can hide or remove sensitive information, but goes
beyond that work by allowing for the possibility of decryption, which supports
richer dataflow structures.

Classical cryptography may be able to support some applications along these
lines.  In particular,
homomorphic encryption enables inference on encrypted
data~\citep{DBLP:journals/corr/XieBFGLN14,DBLP:conf/icml/Gilad-BachrachD16}. On the other hand,
classical cryptographic functions are generally not differentiable, so
they are at odds with training by stochastic gradient descent (SGD),
the main optimization technique for deep neural
networks.  
Therefore, we would have trouble learning \emph{what} to
encrypt, even if we know how to encrypt.  Integrating classical
cryptographic functions---and, more generally, integrating other known
functions and relations
(e.g.,~\citep{DBLP:journals/corr/NeelakantanLS15})---into neural
networks remains a fascinating problem.

Prior work at the intersection of machine learning and cryptography
has focused on the generation and establishment of cryptographic
keys~\citep{DBLP:phd/de/Ruttor2006,kinzel2002neural}, and on
corresponding attacks~\citep{DBLP:conf/asiacrypt/KlimovMS02}. In
contrast, our work takes these keys for granted, and focuses on their
use; a crucial, new element in our work is the reliance on adversarial goals
and training.
More broadly, from the perspective of machine learning, our work relates to the
application of neural networks to multiagent tasks, mentioned
above, and to the vibrant research on generative models and on
adversarial training
(e.g.,~\citep{DBLP:conf/nips/GoodfellowPMXWOCB14,DBLP:journals/corr/DentonCSF15,improvedgan,fgan,infogan,DBLP:journals/corr/GaninUAGLLML15}).
From the perspective of cryptography, it relates to big themes such as
privacy and discrimination.  While we embrace a playful, exploratory
approach, we do so with the hope that it will provide
insights useful for further work on these topics.

Section~\ref{sec:sharedkey} presents our approach to 
learning symmetric encryption (that is, shared-key encryption, in
which the same keys are used for encryption and for decryption) and our corresponding results.
Appendix~\ref{sec:publickey} explains how the same
concepts apply to asymmetric encryption (that is, public-key
encryption, in which different keys are used for encryption and for
decryption).  
Section~\ref{sec:appexperiments} considers selective protection.
Section~\ref{sec:conclusion} concludes and suggests avenues for
further research.
%
% This paper relies only on basic ideas in cryptography,
% and should be accessible to those with modest acquaintance with the subject.
Appendix~\ref{sec:background} is a brief
review of background on neural networks.

% The appreciation of a few secondary points requires more advanced technical knowledge.

\section{Learning Symmetric Encryption}\label{sec:sharedkey}

This section discusses how to protect the confidentiality of plaintexts using shared keys.
It  describes the organization of the system that
we consider, and the objectives of the participants in this system.
It also explains the training of these participants, defines their architecture, 
and presents experiments.

\subsection{System Organization}\label{sec:sharedkeyarchitecture}

A classic scenario in security involves three parties: Alice, Bob, and
Eve.
Typically, Alice and Bob
wish to communicate securely, and Eve wishes to eavesdrop on their
communications. Thus, the desired security property is secrecy (not
integrity), and the adversary is a ``passive attacker'' that can
intercept communications but that is otherwise quite limited: it
cannot initiate sessions, inject messages, or modify messages in
transit.

\begin{figure}
\centering\includegraphics[width=.9\linewidth]{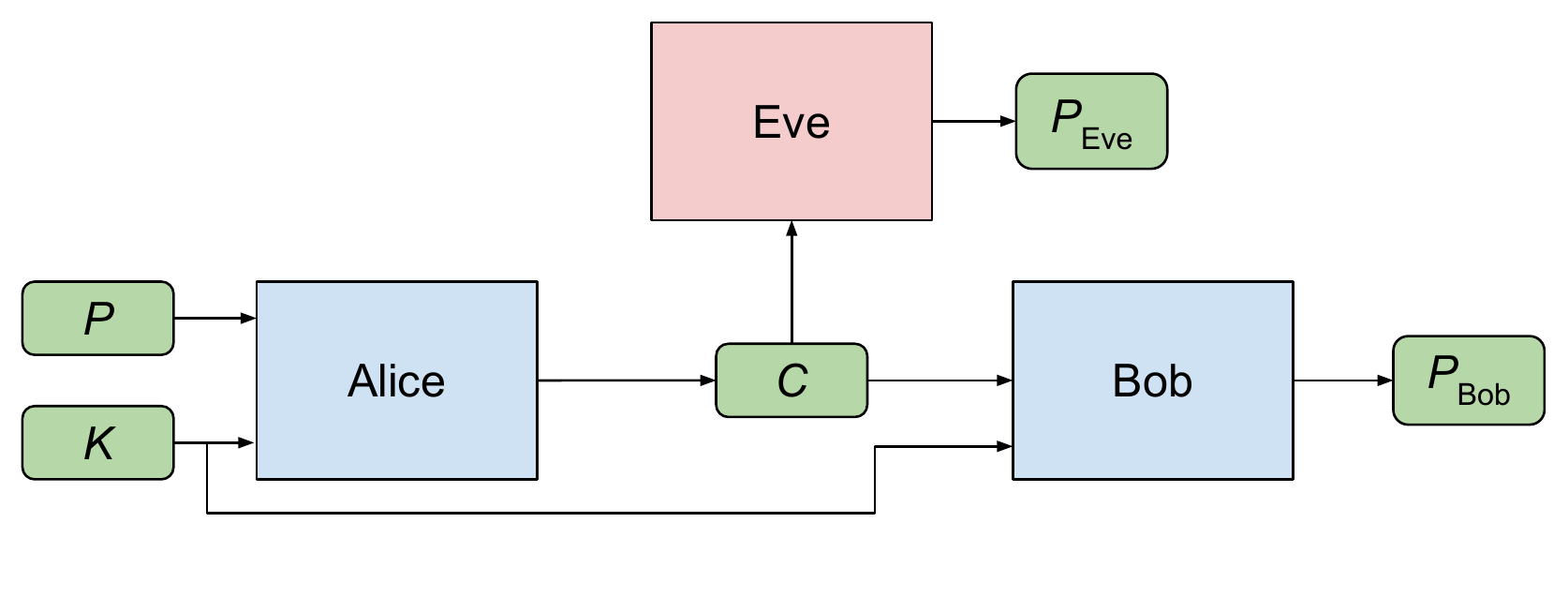}
\caption{Alice, Bob, and Eve, with a symmetric cryptosystem.}\label{fig:symm}
\end{figure}

We start with a particularly simple instance of this scenario,
depicted in Figure~\ref{fig:symm}, in which Alice wishes to send a
single confidential message $P$ to Bob. The message $P$ is an input to
Alice. When Alice processes this input, it produces an output
$C$. (``$P$'' stands for ``plaintext'' and ``$C$'' stands for
``ciphertext''.) Both Bob and Eve receive $C$, process it, and attempt
to recover~$P$.  We represent what they compute by $P_{\mbox{\tiny Bob}}$ and $P_{\mbox{\tiny Eve}}$,
respectively.  Alice and Bob have an advantage over Eve: they share a
secret key $K$. We treat $K$ as an additional input to Alice and Bob.
We assume one fresh key $K$ per plaintext $P$, 
but, at least at this abstract
level, we do not impose that $K$ and~$P$ have the same length.

For us, Alice, Bob, and Eve are all neural networks.
We describe their
structures in Sections~\ref{sec:nnarchitecture} and~\ref{sec:sharedkeyexperiments}.  They each have 
parameters, which we write $\theta_A$, $\theta_B$, and $\theta_E$,
respectively. Since $\theta_A$ and $\theta_B$ need not be equal,
encryption and decryption need not be the same function even if
Alice and Bob have the same structure.
As is common for neural networks, Alice, Bob, and Eve work over tuples
of floating-point numbers, rather than sequences of bits. In other
words, $K$, $P$, $P_{\mbox{\tiny Bob}}$, $P_{\mbox{\tiny Eve}}$, and $C$ are all tuples of floating-point
numbers.
% ; for simplicity, we take these tuples to be all of the same length.
Note that, with this formulation, $C$, $P_{\mbox{\tiny Bob}}$, and $P_{\mbox{\tiny Eve}}$ may consist of arbitrary floating-point
numbers even if $P$ and $K$ consist of 0s and 1s.  In practice, our
implementations constrain these values to the range $(-1, 1)$, but permit
the intermediate values.
We have explored
alternatives (based on Williams' REINFORCE
algorithm~\citep{Williams92simplestatistical} or on Foerster et al.'s discretization technique~\citep{DBLP:journals/corr/FoersterAFW16a}), but omit them as they
are not essential to our main points.

This set-up, although rudimentary, suffices for basic schemes, in
particular allowing for the possibility that Alice and Bob decide to
rely on $K$ as a one-time pad, performing encryption and decryption simply by
XORing the key $K$ with the plaintext $P$ and the ciphertext $C$, respectively.
However, we do not require that Alice and Bob function in this
way---and indeed, in our experiments in
Section~\ref{sec:sharedkeyexperiments}, they discover other schemes.
For simplicity, we ignore the process of generating a key from a seed. We also omit
the use of randomness for probabilistic
encryption~\citep{DBLP:journals/jcss/GoldwasserM84}. Such enhancements
may be the subject of further work.

\subsection{Objectives}\label{sec:objectives}

% Neither Bob nor Eve receive the plaintext $P$, but $P$ is useful for defining their objectives. 

Informally, the objectives of the participants are as follows.
Eve's goal is simple: to reconstruct $P$ accurately (in other words, to minimize the error between $P$ and $P_{\mbox{\tiny Eve}}$).
Alice and Bob want to communicate clearly (to
  minimize the error between $P$ and $P_{\mbox{\tiny Bob}}$), but also to hide their
  communication from Eve.
Note that, in line with modern cryptographic definitions
(e.g.,~\citep{DBLP:journals/jcss/GoldwasserM84}), we do not require
that the ciphertext $C$ ``look random'' to Eve. A ciphertext may even
contain obvious metadata that identifies it as such.  Therefore, it is
not a goal for Eve to distinguish $C$ from a random value drawn from
some distribution. In this respect, Eve's objectives contrast with
common ones for the adversaries of GANs. On the other hand, one could try to
reformulate Eve's goal in terms of distinguishing the 
ciphertexts constructed from two different plaintexts.

Given these objectives, instead of training each of Alice and Bob
separately to implement some known cryptosystem~\citep{dourlensapplied},
we train Alice and Bob jointly to communicate successfully and to defeat Eve without
a pre-specified notion of what cryptosystem they may discover for
this purpose. 
Much as in the definitions of GANs, we
would like Alice and Bob to defeat the best possible version of Eve,
rather than a fixed Eve. 
%MA28: start of introducing keys in losses
Of course, Alice and Bob  may not
win for every plaintext and every key, since knowledge of some particular
plaintexts and keys may be hardwired into Eve. (For instance, Eve could always
output the same plaintext, and be right at least once.) 
Therefore, we 
assume a distribution on plaintexts and keys, and phrase
our goals for Alice and Bob in terms of expected values.

We write $A(\theta_A, P, K)$ for Alice's output on input $P, K$, write $B(\theta_B, C, K)$ for Bob's output on input $C, K$, and write
$E(\theta_E, C)$ for Eve's output on input $C$.
We introduce a distance function $d$ on plaintexts. Although the
  exact choice of this function is probably not crucial, for
  concreteness we take the L1 distance 
$d(P,P') = \Sigma_{i=1,N} |P_i - P'_i|$
where $N$ is the length of plaintexts.
We define a per-example loss function for Eve:
\[
\begin{array}{l}
L_E(\theta_A,\theta_E,P,K) = 
d(P,E(\theta_E, A(\theta_A,P,K)))
\end{array}
\]
Intuitively, $L_E(\theta_A,\theta_E,P,K)$ represents how much Eve is wrong when the plaintext is $P$ and the key is $K$.
We also define a loss function for Eve over the distribution on plaintexts and keys by taking an expected value:
\[
\begin{array}{l}
L_E(\theta_A,\theta_E) =
{\expectedvalue}_{P,K}(d(P,E(\theta_E, A(\theta_A,P,K))))
\end{array}
\]
We obtain the ``optimal Eve'' by minimizing this loss:
\[
O_E(\theta_A) = {\argmin}_{\theta_E}(L_E(\theta_A,\theta_E))
\]
Similarly, we define a per-example reconstruction error for Bob, and extend it
to the distribution on plaintexts and keys:
\[
\begin{array}{l}
L_B(\theta_A,\theta_B,P,K)  =
d(P,B(\theta_B, A(\theta_A,P,K), K))\\[.5em]
L_B(\theta_A,\theta_B) =
{\expectedvalue}_{P,K}(d(P,B(\theta_B, A(\theta_A,P,K),K)))
\end{array}
\]
We define a loss function for Alice and Bob by combining $L_B$ and the optimal value of~$L_E$:
\[
\begin{array}{l}
L_{AB}(\theta_A,\theta_B) =
L_B(\theta_A,\theta_B) - L_E(\theta_A,O_E(\theta_A))
\end{array}
\]
This combination reflects that Alice and Bob want to minimize Bob's
reconstruction error and to maximize the reconstruction error of the
``optimal Eve''.
The use of a simple subtraction is somewhat arbitrary; below we
describe useful variants.
We obtain the ``optimal Alice and Bob'' by 
minimizing $L_{AB}(\theta_A,\theta_B)$:
\[
(O_A,O_B) = {\argmin}_{(\theta_A,\theta_B)}(L_{AB}(\theta_A,\theta_B))
\]

We write ``optimal'' in quotes because there need be no
single global minimum.
In general, there are many equi-optimal
solutions for Alice and Bob.  
As a simple example, assuming that the key is of the same size as the plaintext
and the ciphertext,
Alice and Bob may XOR the plaintext and the ciphertext, respectively, with any permutation of the key, and all permutations are equally good as long as Alice and Bob use the same one; moreover,
with the way we architect our networks (see Section~\ref{sec:nnarchitecture}), 
all permutations are equally likely to arise.

Training begins with the Alice and Bob networks initialized randomly.
The goal of training is to go from that state to $(O_A,O_B)$, or close to 
$(O_A,O_B)$.
We explain the training process next.

\subsection{Training Refinements}\label{sec:sharedkeytraining}

Our training method is based upon SGD.
In practice, much as in work on GANs, our training method cuts a
few corners and incorporates a few improvements with respect to the
high-level description of objectives of
Section~\ref{sec:objectives}. We present these refinements next, and 
give further details in Section~\ref{sec:sharedkeyexperiments}.

First, the training relies on estimated values calculated over ``minibatches'' of hundreds or thousands of examples,
rather than on expected values over a distribution.

We do
not compute the ``optimal Eve'' for a given value of $\theta_A$, but simply approximate it,
alternating the training of Eve with that
of Alice and Bob.
Intuitively, the training may for example proceed roughly as
follows.  Alice may initially produce ciphertexts that neither
Bob nor Eve understand at all. By training for a few steps, Alice and
Bob may discover a way to communicate that allows Bob to decrypt
Alice's ciphertexts at least partly, but which is not understood by
(the present version of) Eve. In particular, Alice and Bob may
discover some trivial transformations, akin to {\tt rot13}. After a
bit of training, however, Eve may start to break this code. With some
more training, Alice and Bob may discover refinements, in particular
codes that exploit the key material better. Eve eventually finds it
impossible to adjust to those codes.
This kind of alternation is typical of games; the theory of continuous
games includes results about convergence to equilibria (e.g.,~\citep{ratliff2013characterization})
which it might be possible to apply in our setting.

Furthermore, in the training of Alice and Bob, we do not attempt to maximize
  Eve's reconstruction error. If we did, and made Eve completely wrong, then Eve
  could be completely right in the next iteration by simply flipping
  all output bits! A more realistic and useful goal for Alice and Bob
  is, generally, to minimize the mutual information between
  Eve's guess and the real plaintext.  In the case of symmetric encryption,
  this goal equates to making Eve produce answers indistinguishable
  from a random guess.
  %% Accordingly, we formulate the loss function
  %% so that it is minimized when half of the message bits are wrong and
  %% half are right.
  This approach is somewhat analogous to methods that aim to prevent
  overtraining GANs on the current adversary~\citep[Section~3.1]{improvedgan}.
Additionally, we can tweak the loss functions so that they do not give 
  much importance to Eve being a little lucky or to Bob making small errors that 
  standard error-correction could easily address.

Finally, once we stop training Alice and Bob, and they have picked
  their cryptosystem, we validate that they work as intended by
  training many instances of Eve that attempt to break the cryptosystem.
  Some of these instances may be derived from earlier
  phases in the training.

\subsection{Neural Network Architecture}\label{sec:nnarchitecture}

%% \begin{figure*}
%% \centering\includegraphics[width=.8\linewidth]{conv}
%% \caption{Neural network layout (sketch).}\label{fig:conv}
%% \end{figure*}

\paragraph{The Architecture of Alice, Bob, and Eve}
Because we wish to explore whether a general neural network can learn to communicate securely, rather than to 
engineer a particular method, 
we aimed to create a neural network
architecture that was \emph{sufficient} to learn mixing
functions such as XOR, but that did not strongly encode the
form of any particular algorithm.
%MA: ``for doing so'' was a bit unclear imho.

To this end, we chose the following ``mix \& transform''  architecture. It has a first fully-connected (FC)
layer, where the number of outputs is equal to the number of
inputs.  The plaintext and key bits are fed into this FC layer.
Because each output bit can be a linear combination of all of
the input bits, this layer 
enables---but does not mandate---mixing between the key and the
plaintext bits. 
%MA: ``doing so or how to do so'' seemed poor, imho
In particular, this layer can permute the bits.
The FC layer is followed by a sequence of convolutional layers,
the last of which produces an output of a size suitable for a plaintext or ciphertext.
These convolutional layers learn to apply
some function to groups of the bits mixed by the previous layer,
without an a priori specification of what that function should be.  
%MA: I did not like the sentence on XOR, and it seemed repetitive anyway, so I omitted it
% (Figure~\ref{fig:conv})
%
Notably, the opposite order (convolutional followed by FC)
is much more common in image-processing
applications.  Neural networks developed for those applications frequently
use convolutions to take advantage of spatial locality.
For neural cryptography, we specifically wanted locality---i.e., which
bits to combine---to be a \emph{learned} property, instead of
a pre-specified one.  While it would certainly work to
manually pair each input plaintext bit with a corresponding
key bit, we felt that doing so would be uninteresting.

We refrain from imposing further constraints that would
simplify the problem.  For
example, we do not tie the parameters $\theta_A$ and $\theta_B$, as
we would if we had in mind that Alice and Bob should both learn the
same function, such as XOR.

\subsection{Experiments}\label{sec:sharedkeyexperiments}

As a proof-of-concept, we implemented Alice, Bob, and Eve networks that take $N$-bit random plaintext and key values, and produce $N$-entry floating-point ciphertexts, for $N = 16$, $32$, and~$64$. Both plaintext and key values are uniformly distributed. Keys are not deliberately reused, but may reoccur because of random selection. (The experiments in Section~\ref{sec:appexperiments} consider more interesting distributions and also allow plaintext and key values to have different sizes.) 

We implemented our experiments in
% Google's open-source machine learning infrastructure,
TensorFlow~\citep{DBLP:journals/corr/AbadiABBCCCDDDG16,DBLP:journals/corr/AbadiBCCDDDGIIK16}.
We ran them on a workstation with one GPU;
the specific computation platform does not affect the outcome of
the experiments.  We plan to release the source code for the
experiments.

\paragraph{Network Structure}

Our networks follow the ``mix \& transform'' pattern described in Section~\ref{sec:nnarchitecture}.
The Alice network concatenates
two $N$-bit inputs (the plaintext and the key) into a $2N$-entry vector,
using $-1$ and $1$ to represent the bit values.  This vector is processed through
a $2N \times 2N$ FC layer, then sent through a succession of four
1-D convolutional layers.  The convolutional layers are described in terms of their
window size, input depth, and output depth.  Each has a ``stride''---the amount
by which the window is shifted at each step. The concrete layers are 
$[4, 1, 2]$, $[2, 2, 4]$, $[1, 4, 4]$, and $[1, 4, 1]$, with strides of $1, 2, 1, 1$.
Intuitively, the first layer slides a size-4 window across the $2N$ output elements
from the FC layer, outputting two numbers (output depth 2), with stride $1$.  We use a
sigmoid nonlinear unit after each layer except the final one.  After the final layer, where
the output is reduced to $N$ elements, we use a tanh nonlinear unit.  (Recall that tanh 
has an output domain between $[-1, 1]$, bringing the values back to a range that can map to
binary values.)
The Bob network is identical to the Alice network
%MA: It is just identical, since ciphertexts and plaintexts have the same shape, so I cut
% ``except that instead of a \{plaintext, key\} input, it receives a \{ciphertext, key\}'' input.
The Eve network takes only the ciphertext as input, and thus has a first
$N \times 2N$ FC layer.

\paragraph{Training}
To train the networks, we use a ``minibatch'' size ranging from 
256 to 4096 entries.  Unless otherwise specified, we present
the results from the training runs with 4096 entries.
We use TensorFlow's Adam~\citep{DBLP:journals/corr/KingmaB14} optimizer 
with a learning rate of $0.0008$.
We do not reduce the learning rate over time because we wish 
Alice, Bob, and Eve to be able to continue to respond strongly 
to changes in other components, until Alice and Bob---hopefully---achieve a solution that is
robust to additional changes in Eve.

Training alternates between Alice/Bob and Eve, with Alice/Bob training for one
minibatch, and then Eve training for two minibatches.  We chose this ratio
in order to give a slight computational edge to the adversary Eve without training it so
much that it becomes excessively specific to the exact current parameters of Alice
and Bob.  Empirically, this value has worked well.

\paragraph{Loss Functions}

Eve's loss function is exactly as described above:  the L1 distance
between Eve's guess and the input plaintext.
The loss function for Alice and Bob is more complex, as indicated in Sections~\ref{sec:objectives} and~\ref{sec:sharedkeytraining}. 
This function has two components, related to Bob's reconstruction error and to the eavesdropper's success.
The first component is 
simply the L1 distance
between Bob's output and the input plaintext.  The latter 
component, on the other hand, is
% \[\frac{\left(N/2 - \mbox{Eve L1 error}\right)^{2}}{(\frac{N}{2})^2}\]
${\left(N/2 - \mbox{Eve L1 error}\right)^{2}}/{({N}/{2})^2}$.
This definition expresses the goal, described in
Section~\ref{sec:sharedkeytraining}, that Eve should not do better
than random guessing.
Accordingly, this component is minimized when half of the message bits
are wrong and half are right. We choose a quadratic formula in order
to place more emphasis on making Eve have a large error, and to impose
less of a penalty when Eve guesses a few bits correctly, as should
happen occasionally even if Eve's guesses are effectively random.
Adopting this formulation allowed us to have a meaningful per-example
loss function (instead of looking at larger batch statistics), and
improved the robustness of training. Its cost is that our final,
trained Alice and Bob typically allow Eve to reconstruct slightly more
bits than purely random guessing would achieve.
We have not obtained satisfactory results for loss functions that
depend linearly (rather than quadratically) on Eve's reconstruction error.
The best formulation remains an open question.

\paragraph{Post-training Evaluation}

After successfully training the networks so that they cross an accuracy threshold
(e.g., at most 0.05 bits of reconstruction error for Alice and Bob, with 
Eve achieving only 1-2 bits more than random guessing would predict),
we reset the Eve network and train it from scratch 5 times, each for up to 250,000
steps, recording the best result achieved by any Eve.  An Alice/Bob combination that fails
to achieve the target thresholds within 150,000 steps is a training
failure.  If the retrained Eves obtain a substantial advantage, the solution
is non-robust.  Otherwise, we consider it a successful training outcome.

\paragraph{Results}

\begin{figure}
\centering\includegraphics[width=.9\linewidth]{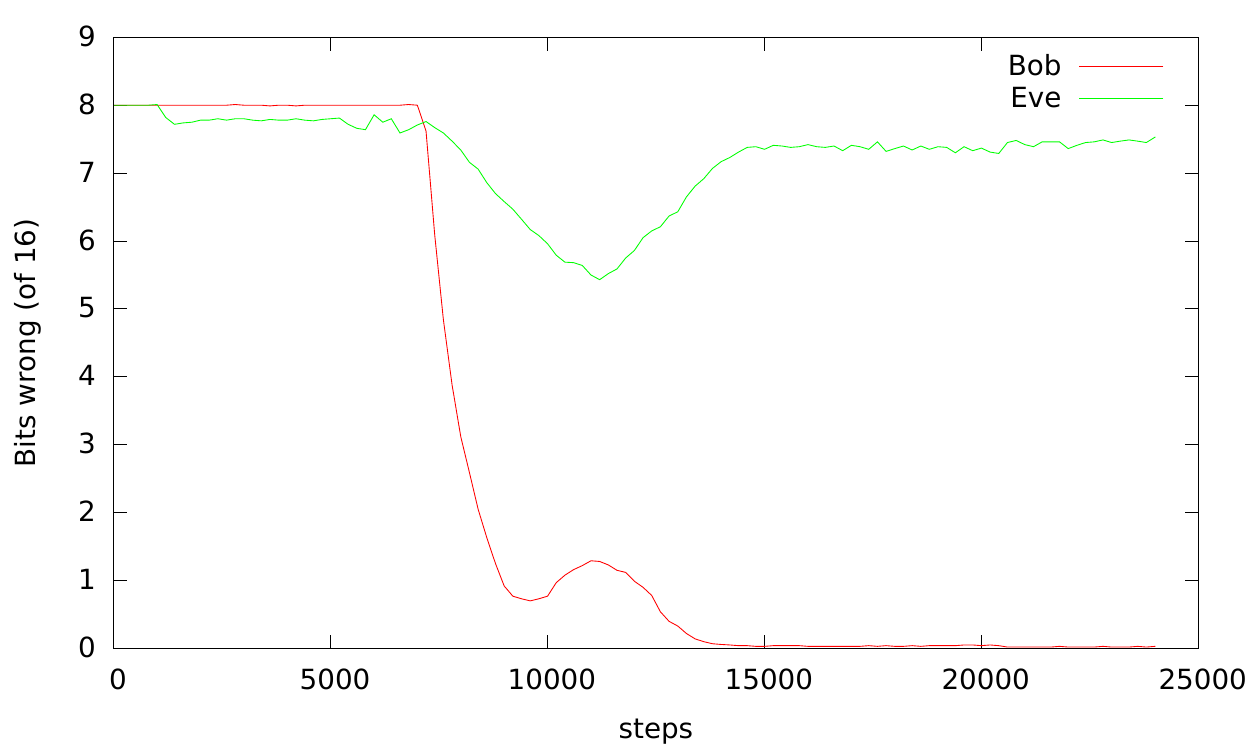}
\caption{Evolution of Bob's and Eve's reconstruction errors during training.  Lines represent the mean error across a minibatch size of 4096.}\label{fig:bob_vs_eve_one}
\end{figure}

Figure~\ref{fig:bob_vs_eve_one} shows, for one successful run, the evolution of 
Bob's reconstruction error 
and Eve's reconstruction error vs.~the number of training steps for $N=16$
bit plaintext and key values, using a minibatch size of 4096.
Each point in the graph is the mean error across 4096 examples.
An ideal result would have Bob's
reconstruction error drop to zero and Eve's reconstruction error reach
8 (half the bits wrong).  In this example, both reconstruction errors
start high.
After a period of time, Alice and Bob start to communicate quite effectively,
but in a way that allows Eve to improve its understanding as well, gradually.
Then, around step 10,000, Alice and Bob counter Eve's progress.
By about step 15,000, the training goals are effectively achieved.  The remaining steps merely increase Eve's reconstruction error slightly.
%MA: I rewrote and simplified above, both because it seemed a little wrong (potentially) and a little repetitive of things said earlier.

\begin{figure}
\centering\includegraphics[width=.9\linewidth]{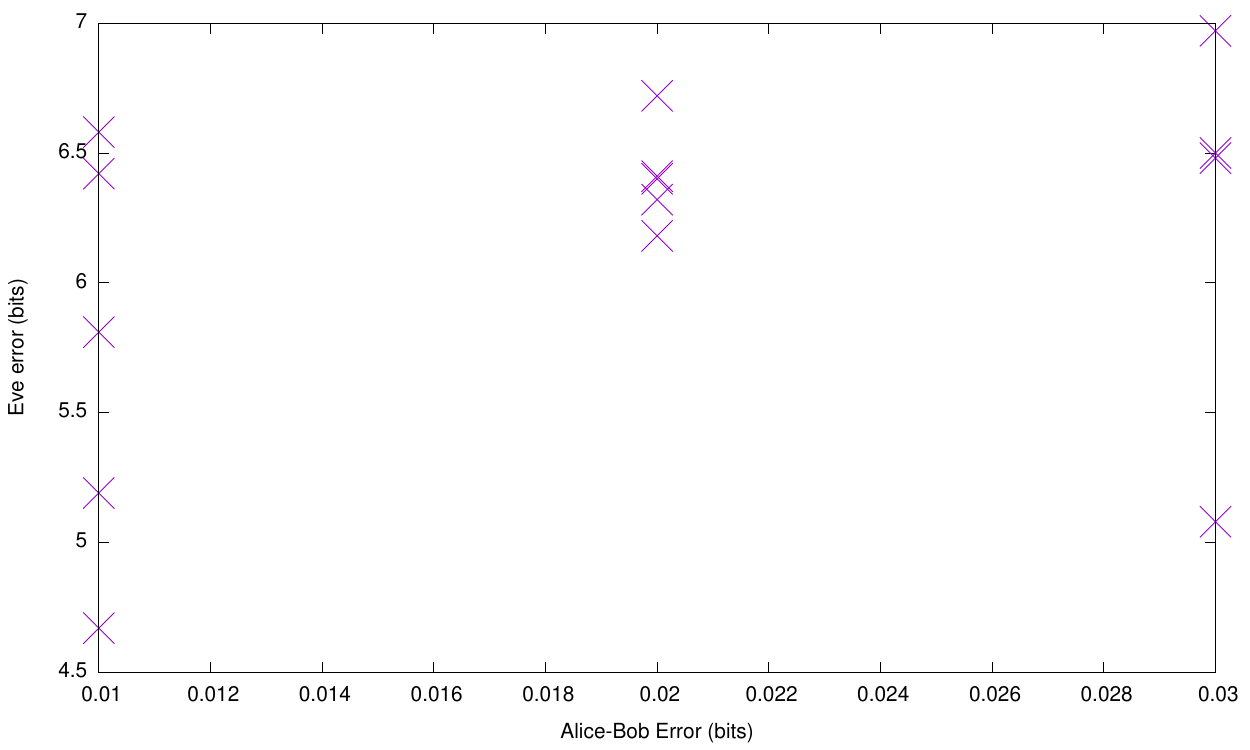}
\caption{Final reconstruction errors of Bob and of the most effective retrained Eve, for the fourteen initially successful runs, of twenty.
An ideal result would be a dot in the upper-left corner, representing no Bob reconstruction error and 8 bits of Eve reconstruction error.}\label{fig:bob_vs_eve_two}
\end{figure}

This training graph does \emph{not} look like a typical result for neural networks, where
monotonicity in the number of steps is generally expected.
%MA29 ``adversarial training'' is an abused phrase in ML, so I added ``this'' to differentiate from other kinds
Instead, the dynamics of this adversarial training appear somewhat more reminiscent
of evolutionary processes.  These dynamics appear to depend on somewhat random-seeming
change to cause the bits to mix slightly, but once there is some mixing,
the gradient descent can rapidly drive it farther.  

Supporting this interpretation is the observation that training is not always successful.
With $N=16$, 
six of twenty initial runs 
were failures that never got Bob's reconstruction error under the
0.05 threshold, or failed to drive Eve's reconstruction error above 7.3 bits (of 16).
%MA29: This used to say
%    The remaining 14 were ``successes``.
% but I would rather not use the ``success'' term too much if it may
% be unclear whether it is a final success or an intermediate success
% that is not necessarily robust.
In order to test the robustness of the other fourteen Alice/Bob combinations,
we retrained Eve five times, and obtained  reconstruction errors for Eve that
ranged from 4.67 to 6.97 bits, with a mean of 6.1.
Figure~\ref{fig:bob_vs_eve_two} shows the final reconstruction errors of Bob and of the most effective retrained Eve
for those fourteen Alice/Bob combinations.
If we somewhat arbitrarily
define success as maintaining Bob's reconstruction error at or under 0.05 bits, and requiring that Eve get at least 6 bits wrong, on average, then
training succeeded half of the time (ten of twenty cases).

%MA29: ``unstable'' is the simplest term that they use, but the literature does not seem to break down all the ways
%      in which there is instability. There are nice things that one could say about Nash equilibria
%      of non-convex games, perhaps, but that may require much more thinking.
Although training with an adversary is often unstable~\citep{improvedgan},
we suspect that some additional engineering of the neural network and 
its training may be able to 
increase this overall success rate.  With a minibatch size of only 512, for example, we achieved a success rate of only ${1}/{3}$ (vs.~the ${1}/{2}$ that we achieved with a minibatch size of 4096).
In the future, it may be worth studying the impact of minibatch sizes, and
also that of other parameters such as the learning rate.

Analogous results hold in general for $N=32$ and $N=64$-bit keys and plaintexts; training appears to be successful somewhat more often for $N=64$.
Basically, the experiments for $N=32$ and $N=64$ indicate that there is nothing special about $N=16$ which, to a cryptographer, may look suspiciously tiny.
We focus our presentation on the case of $N=16$ because, first, the experiments
run more rapidly, and second, it is modestly easier to examine their
behavior. 

For one successful training run, we studied the changes in the ciphertext
induced by various plaintext/key pairs. Although we did not perform 
an exhaustive
analysis of the encryption method, we did make a few observations.
First, it is key-dependent:  changing the key and holding the plaintext constant results in different ciphertext output. It
is also plaintext-dependent, as required for successful communication.
However, it is not simply XOR. In particular, the output values are 
often floating-point values other than 0 and 1.
Moreover, the effect of a change to either a key bit or a plaintext bit is spread across multiple elements in the ciphertext,
      not constrained to a single bit as it would be with XOR.  
      A single-bit flip in the key 
      typically induces significant changes in three to six 
      of the 16 elements in the ciphertext, and smaller changes in other elements.
      Plaintext bits are similarly diffused across the ciphertext.

%% \begin{itemize}
%% \item It \emph{is} key-dependent:  changing the key and holding the plaintext constant results in different ciphertext output.
%% \item It \emph{is} plaintext-dependent, as required for successful communication.
%% \item It is not simply XOR. In particular, the output values are 
%% often floating-point values other than 0 and 1.
%% \item The effect of a change to either a key bit or a plaintext bit is spread across multiple elements in the ciphertext,
%%       not constrained to a single bit as it would be with XOR.  
%%       A single-bit flip in the key 
%%       typically induces significant changes in three to six 
%%       of the 16 elements in the ciphertext, and smaller changes in other elements.
%%       Message bits are similarly diffused across the ciphertext.
%% \end{itemize}

\section{Learning What to Protect}\label{sec:appexperiments}

Building on the results of Section~\ref{sec:sharedkey}, we study
selective protection. In other words, we consider the question of
whether neural networks can learn what information to protect, given
confidentiality objectives described in terms of an adversary.  In the
simplest case, a plaintext may have several components; if we wish to
keep the adversary from seeing one of the components, it may
suffice to encrypt it. More generally, we may wish to publish some values correlated with the plaintext but
prevent the adversary from calculating other values.
In such scenarios, protecting information selectively while
maximizing utility is both challenging and interesting. 

\subsection{The Problem}

To test these ideas and to demonstrate the viability of selective protection,
we focus on a concrete experiment. 
We construct an artificial dataset consisting
of tuples of four values, $\left<\mbox{A, B, C, D}\right>$.  
We aim to build and train a system that outputs two
predictions of D, given the first three values as inputs:
  a ``true'' prediction of D (that is, the most accurate possible
estimate of~D given $\left<\mbox{A, B, C}\right>$), and
  a ``privacy-preserving'' estimate of D, which we call D-public,
defined as the best possible estimate of D that does not reveal any
information about the value of~C.  

The system organization is essentially that of Figure~\ref{fig:symm},
with Alice and Bob sharing a key, but here Alice receives
$\left<\mbox{A, B, C}\right>$ and produces D-public in addition to a
ciphertext; both Bob and Eve have access to Alice's outputs; Bob uses
them for producing an improved estimate of D, while Eve attempts to
recover C.  Thus, our goal is to demonstrate that the adversarial
training permits approximating D without revealing C, and that this
approximation can be combined with encrypted information and with a
key in order to obtain a better approximation of~D.

The distributions of A, B, C, and D are constructed
so that there is substantial correlation between them. In particular, 
knowing D exactly would convey some information about the likely values
of~C.  In our experiments, A, B, C, and D are normally distributed
floating-point values with a correlation matrix imposed upon them.
The specifics of how we generate A, B, C, and D are standard: we compute the Cholesky decomposition $L L^*$ of a 4 $\times$ 4 correlation matrix that specifies the pairwise correlations, we generate vectors of 4 normally distributed floating-point values, and then we multiply these vectors by $L$ in order to generate values of A, B, C, and D.

Although A, B, C, and D do not correspond to any real-world
data, these experiments are loosely inspired by real-world
tasks. For example, A,
B, and C may represent attributes of a user, which may in general be
correlated, and D may represent a decision about the user, a
prediction of the user's behavior, or a recommendation to the user.

\subsection{Neural Network Architecture and Training Goals}
In the experiments that follow, we use an augmented version of the neural network architecture of Section~\ref{sec:nnarchitecture}.
The inputs first go into a new FC layer (12 inputs---eight key bits and four values---and 12 outputs);  the outputs of that first layer are fed into a network with the architecture 
of Section~\ref{sec:nnarchitecture}.
Intuitively, we chose this augmented architecture because a single
FC layer should be capable of predicting D from A, B, and
C, as well as making a prediction decorrelated with C; and the
architecture of Section~\ref{sec:nnarchitecture} suffices to encrypt
any of the output of the first layer under the key.  We therefore
believed this augmented architecture would be sufficient to accomplish
its task, though it may be more than is necessary to do so.

We train Eve with the goal of minimizing the squared error
of its estimate  of C.
We train Alice and Bob with the goal of minimizing a linear combination of three terms:
the squared error of each of the two estimates of D, and the absolute value of the covariance of Eve's estimate of C
with respect to the actual value of~C.  We compute this covariance on a batch of training
examples, and define the loss function for Alice and Bob batch-wise.

A human engineer would naturally try to design the system somewhat
differently---likely first computing the desired quantities, and then
encrypting the values to be hidden, relying on the encryption and
decryption components developed in Section~\ref{sec:sharedkey} as
modules.  
We prefer instead an end-to-end approach, because our goal
is to understand whether a neural network can learn to hide
information selectively, given only a high-level specification
(``do not leak information about C, but let people with the right key
learn as much as possible about D'').  In this end-to-end approach,
the system does not have distinct, pre-assigned units for encryption
and decryption, and all components are trained jointly. In particular,
although we benefit from some of the insights from
Section~\ref{sec:sharedkey}, we do not rely on the components
developed there.
End-to-end approaches are sometimes preferred in machine learning
because they can more efficiently use resources and
information. In our case, this feature may mean, in particular, that
the cryptographic operations are optimized for the given distribution
of inputs. Nevertheless, our choices have the primary goal of maximizing
insight, not efficiency.

\subsection{Results}

To understand whether the system is learning to hide information
properly, we train a separate evaluator that we call ``Blind Eve'',
which is aware of the distribution of~C.  Blind Eve tries to
guess C relying only upon this baseline information, whereas the
real Eve also knows the intermediate output (D-public) and the ciphertext.
If Eve's reconstruction error becomes equal to that
of Blind Eve, we know that Eve is not successfully extracting
information from the public estimate and the ciphertext.

\begin{figure}
\centering\includegraphics[width=.9\linewidth]{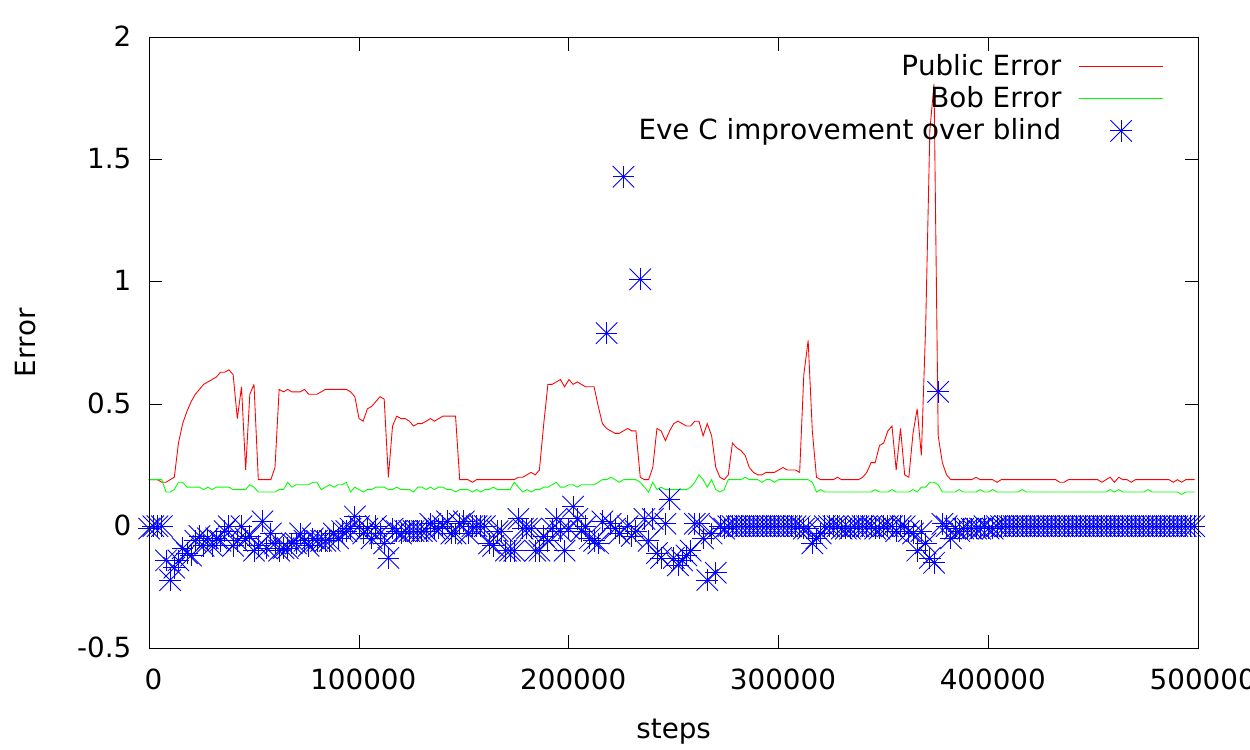}
\caption{Training to estimate D while hiding C.}\label{fig:hide_c}
\end{figure}

Figure~\ref{fig:hide_c} shows the results of adversarial training to learn
to hide C.  The red ``public error'' line shows the average error
of D-public vs. D.  The green ``Bob error'' line shows Bob's average
reconstruction error, which drops over time until Bob is making the best possible prediction of D given the input.
The public error becomes low, but is slightly higher than
Bob's.  The blue dots show Eve's average error \emph{improvement} over
Blind Eve.  After a few steps of training, Eve's error is substantially lower than that of Blind Eve (by 0.1--0.2).  As time proceeds,
Eve's relative advantage becomes zero.
Eve is not able to reconstruct any
more information about C than would be possible by simply knowing the 
distribution of values of C.

%% As an another example, A, B, and C may be
%% attributes of an object, say a bottle of wine or a piece of candy, and
%% D an estimate of its desirability. We may decide to hide some of the
%% attributes in calculating this desirability, perhaps with the goal of
%% protecting against misleading biases in the training data.

\section{Conclusion}\label{sec:conclusion}

In this paper, we demonstrate that neural networks can learn to
protect communications. The learning does not require prescribing a
particular set of cryptographic algorithms, nor indicating ways of
applying these algorithms: it is based only on a secrecy specification
represented by the training objectives. In this setting, we model
attackers by neural networks;
alternative models may perhaps be enabled by
reinforcement learning.
% while obviously limited, this class of attackers may still be appropriate for certain applications with confidentiality requirements.

There is more to cryptography than encryption.  In this spirit, further
work may consider other tasks, for example steganography, pseudorandom-number
generation, or integrity checks. Finally, neural networks may be useful not only for cryptographic protections
but also for attacks. While it seems improbable that neural
networks would become great at cryptanalysis, they may be quite
effective in making sense of metadata and in traffic analysis.

\subsubsection*{Acknowledgments}
We are grateful to
Samy Bengio,
Laura Downs,
\'Ulfar Erlingsson,
Jakob Foerster,
Nando de Freitas,
Ian Goodfellow,
Geoff Hinton,
Chris Olah,
Ananth Raghunathan,
and Luke Vilnis
for discussions on the matter of this paper.

\bibliographystyle{iclr2017_conference}
% \bibliography{neurocrypt}

\newpage

\appendix

\section*{Appendix}

\section{Learning Asymmetric Encryption}\label{sec:publickey}

\begin{figure}
\centering\includegraphics[width=.9\linewidth]{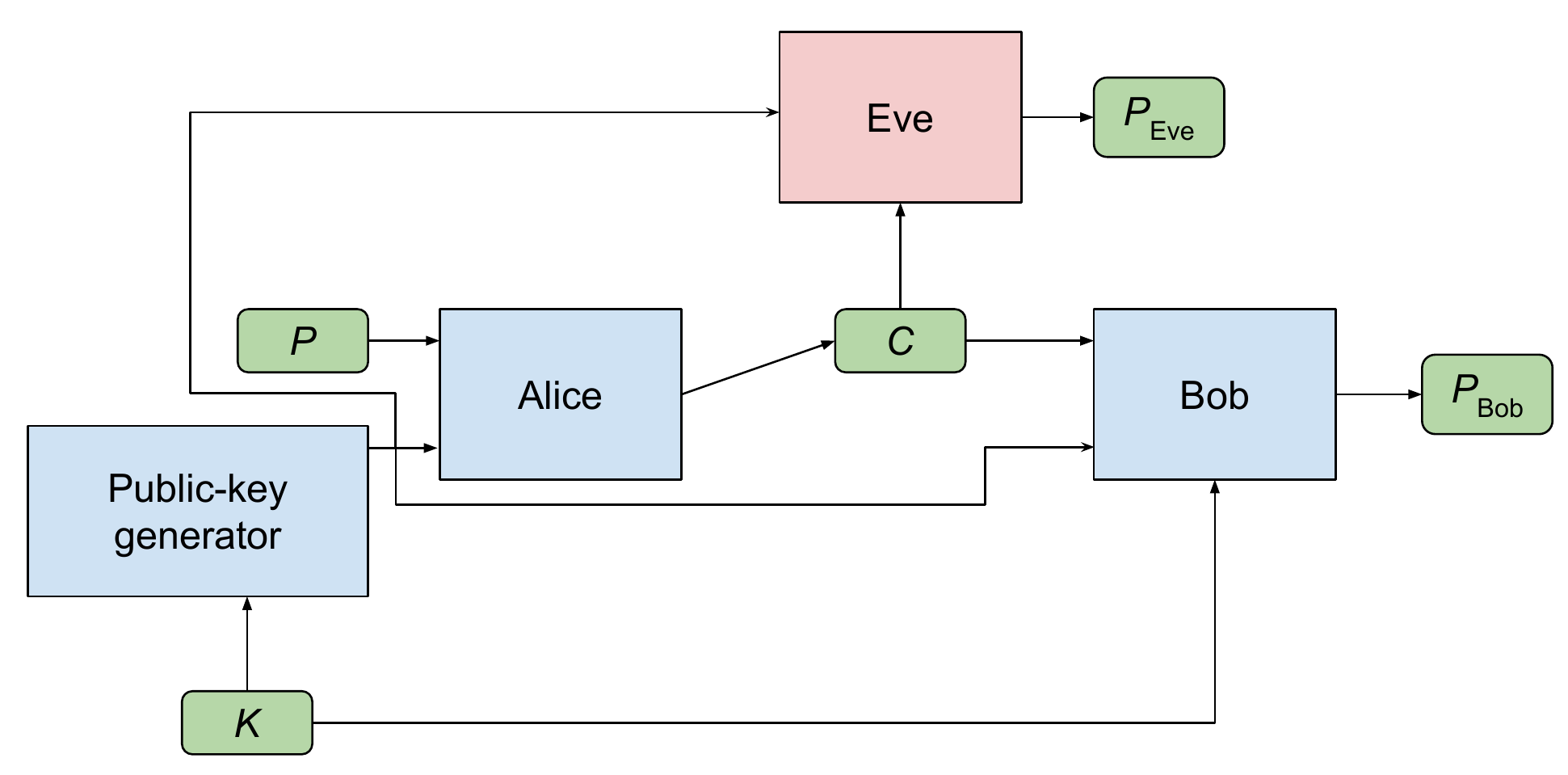}
\caption{Alice, Bob, and Eve, with an asymmetric cryptosystem.}\label{fig:asymm}
\end{figure}

Paralleling Section~\ref{sec:sharedkey}, this section examines
asymmetric encryption (also known as public-key encryption).  It
presents definitions and experimental results, but 
omits a detailed discussion
of the objectives of asymmetric encryption, of the corresponding loss
functions, and of the practical refinements that we develop for
training, which are analogous to
those for symmetric encryption.

\subsection{Definitions}\label{sec:publickeyarchitecture}

In asymmetric encryption, a
secret is associated with each principal. The secret may be seen as a
seed for generating cryptographic keys, or directly as a secret key;
we adopt the latter view. A public key can be derived from the secret,
in such a way that messages encrypted under the public key can be
decrypted only with knowledge of the secret.

We specify asymmetric encryption using a twist on our specification for
symmetric encryption, shown in Figure~\ref{fig:asymm}.  Instead of directly
supplying the secret encryption key to Alice, we supply
the secret key to a public-key generator, the output of which is available
to every node.  Only Bob has access to the underlying secret key.
Much as in Section~\ref{sec:sharedkey}, several variants are possible,
for instance to support probabilistic encryption.

The public-key generator is itself a neural network, with its own
parameters.  The loss functions treats these parameters much like
those of Alice and Bob. In training, these parameters are adjusted at
the same time as those of Alice and Bob.

\subsection{Experiments}\label{sec:publickeyexperiments}

%MA28: felt that a little more needed to be said to start this section
In our experiments on asymmetric encryption, we rely on the same approach 
as in Section~\ref{sec:sharedkeyexperiments}. In particular, we adopt the 
same network structure and the same approach to training.

The results of these experiments are intriguing, but much harder
to interpret than those for symmetric encryption.  In most training runs,
the networks failed to achieve a robust outcome.
Often, although it appeared
that Alice and Bob had learned to communicate secretly, upon resetting and retraining Eve, the retrained adversary was able to decrypt messages nearly as well as Bob was.

\begin{figure}
\centering\includegraphics[width=.9\linewidth]{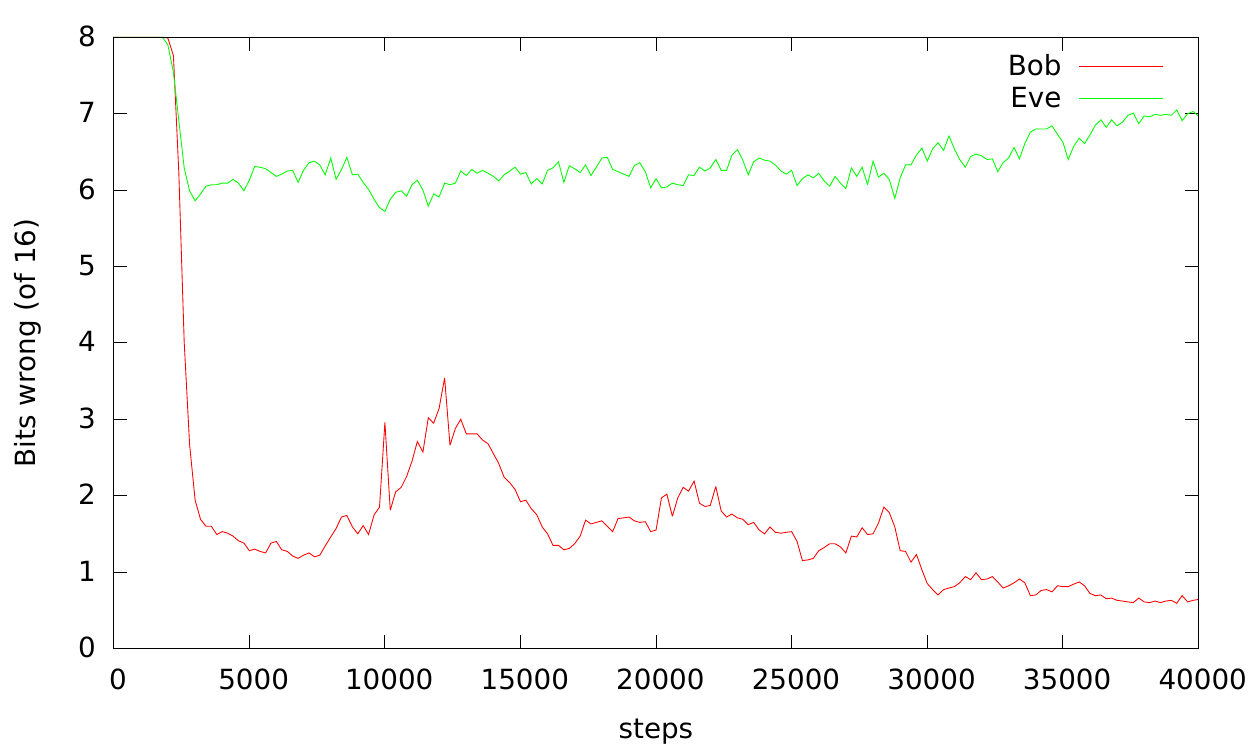}
\caption{Bob's and Eve's reconstruction errors with an asymmetric formulation.}\label{fig:asymm-result}
\end{figure}

However, Figure~\ref{fig:asymm-result} shows
the results of \emph{one} training run, in which even after five reset/retrain cycles,
Eve was unable to decrypt messages between Alice and Bob.

Our chosen network structure is not sufficient to learn general
implementations of many of the mathematical concepts underlying modern asymmetric
cryptography, such as integer modular arithmetic.  We therefore believe that the
most likely explanation for this successful training run was that Alice and Bob
accidentally obtained some ``security by obscurity''
(cf.~the derivation of asymmetric schemes from symmetric schemes by obfuscation~\citep{Barak:2012:POP:2160158.2160159}).  This
belief is somewhat reinforced by the fact that the training result was
fragile:  upon further training
of Alice and Bob, Eve \emph{was} able to decrypt the messages.
However, we cannot rule out that the networks trained into some
set of hard-to-invert matrix operations resulting in ``public-key-like''
behavior.  Our results suggest that this issue deserves more exploration.

Further work might
attempt to strengthen these results, perhaps relying on new
designs of neural networks or new training procedures. A modest next step may
consist in trying to learn particular asymmetric algorithms, such as
lattice-based ciphers, in order to identify the required
neural network structure and capacity.

\section{Background on Neural Networks}\label{sec:background}

Most of this paper assumes only a few basic notions
in machine learning and neural networks, as provided by general
introductions (e.g.,~\cite{deeplearning}). The following is a brief review.

Neural networks are specifications of parameterized functions.
They are typically constructed out of a sequence of somewhat
modular building blocks.  For example, the input to Alice
is a vector of bits that represents the concatenation
of the key and the plaintext.  This vector ($x$) is input into a ``fully-connected''
layer, which consists of a matrix multiply (by $A$) and a vector addition (with $b$):  
$Ax+b$.  The result of that operation is then passed into a nonlinear function,
sometimes termed an ``activation function'', such as the sigmoid function,
or the hyperbolic tangent function, tanh.  In classical neural networks,
the activation function represents a threshold
that determines whether a neuron would ``fire'' or not, based upon its inputs.
This threshold, and matrices and vectors such as $A$ and $b$, are
typical neural network ``parameters''.
``Training'' a neural network is the process that finds values
of its parameters that minimize the specified
loss function over the training inputs.

Fully-connected layers are powerful but require substantial amounts
of memory for a large network.
An alternative to  fully-connected layers are ``convolutional'' layers.
Convolutional layers operate much like their
counterparts in computer graphics, by sliding a parameterized
convolution window across their input.  
The number of parameters in this window is much smaller than
in an equivalent fully-connected layer.
Convolutional layers are useful for applying
the same function(s) at every point in an input.

A neural network architecture consists of a graph of these building
blocks (often, but not always, a DAG), specifying what
the individual layers are (e.g., fully-connected or convolutional),
how they are parameterized (number of inputs, number of
outputs, etc.), and how they are wired.

\end{document}